# PERFORMANCE ANALYSIS OF THE NEWLY DEVELOPED OPERATIONAL AMPLIFIER ARD820

Aigars Atvars[1], Dmitry Kostrichkin[2], Sergey Rudenko[2], Mihails Lapkis[2]

[2]University of Latvia, Latvia; [2]JSC RD ALFA Microelectronics, Latvia

Aigars.Atvars@lu.lv, D.Kostrichkin@rdalfa.lv, Rudenko@rdalfa.lv, Mihails.Lapkis@rdalfa.lv

**Abstract.** Operational amplifiers are critical components in modern analog electronics, widely used across applications such as signal conditioning, instrumentation, voltage regulation, and analog-to-digital conversion. This paper presents the design, development, and evaluation of the aRD820, a low-power, rail-to-rail operational amplifier, as a cost-effective alternative to the industry-standard AD820. The aRD820 is optimized for operation on a single supply voltage ranging from 5 to 30 V or a dual supply of ±2.5 V to ±15 V. Key design objectives included achieving low voltage noise (< 4 µV peak-to-peak from 0.1 to 10 Hz), ultra-low input bias current (< 15 pA), and a low offset voltage (< 500 µV), making it suitable for high-sensitivity applications. In adapting the AD820 design, RD Alfa Microelectronics addressed specific production constraints by modifying key circuit elements to meet the specifics of available production facilities. The modified input stage reduced noise by incorporating source followers, enhancing the allowable input signal range. Adjustments to the second and output stages improved phase stability and gain characteristics. Extensive testing of the aRD820 in both wafer form and TO-5 packaged chips confirmed that the amplifier meets or exceeds planned specifications. Comparative results demonstrate that the aRD820 provides open-loop gain, offset voltage stability, and output saturation voltages comparable to those of the AD820. Noise measurements yielded values within the planned range, with an input noise density of approximately 13.5–18.1 nV/√Hz, suitable for applications requiring precision signal processing. This paper provides a comprehensive overview of the design modifications and test results, establishing the aRD820 as a viable alternative to existing op-amps in similar applications.

## Introduction

Operational amplifiers (op amps) are essential components in analog electronics [1]. They can be used in differential DC amplifiers, analog integrators, voltage regulators, voltage-to-current converters, active filters, analogue-to-digital and digital-to-analogue converters, peak detectors, waveform generators, etc. [2,3]. Op amps typically offer high gain (ranging from 100,000 to 1,000,000 V/V), high input impedance (which can range from 1 MΩ to around 10 GΩ), and low output impedance (generally below 100 Ω), making them adaptable for various circuit configurations. Unlike traditional amplifiers such as audio power amplifiers, which are designed to provide significant output power (usually between 10W to 100W or more) to drive loads like speakers, op-amps are engineered for precision and adaptability. They are often used as building blocks for complex analog functions and typically produce lower power output, in the range of 10 mW to 100 mW, depending on the specific application and supply voltage.

Operational amplifiers can be broadly classified into several types, each designed to meet specific requirements. There are operational amplifiers with accent on high speed (bandwidth > 50 MHz) [4], precision (offset voltage < 1 mV), low input bias current (< 100 pA), low power (< 1 mA/amp), low noise (<2 µV), rail-to-rail [5, 6], high output current (> 100 mA), or high voltage (Vs > 12 V). Voltage Operational Amplifiers amplify input voltage signals, essential for applications where stable voltage gain is needed. Current Operational Amplifiers, also known as transimpedance amplifiers, amplify input current signals and convert them to an output voltage. Analog Operational Amplifiers are designed to handle continuous-time signals. Digital Operational Amplifiers are designed to work in mixed-signal environments, where both analog and digital signals coexist. Analog operational amplifiers are generally more popular, simpler, more cost-effective, and ideal for a broad range of general-purpose amplification needs, whereas digital op-amps are often used in specialized applications requiring programmability and digital control. For low signal detection, only low-noise analog operational amplifiers can be used.

The constant aims of op-amp developers are to minimize their noise [7], maximize the power supply rejection ratio [8], make them robust to electromagnetic backgrounds [9,10], and ensure their long-term performance [11]. Several methods have been developed to boost the performance of op amp, e.g., adaptive biasing for CMOS op amp [12], the use of recycling folded cascodes for operational transconductance amplifiers (OTA) [13], employment of planar Complementary Metal-Oxide-Semiconductor (CMOS) and multigate transistors for heavy resistance load OTA buffer amplifiers [14], flexible noise-power balancing schemes for CMOS op-amps [15], and achievement of high performance op amps using coplanar amorphous indium–gallium–zinc oxide thin-film transistors [16].

During the design phase, op amps are typically modelled by specialized software like SPICE (Simulation Program with Integrated Circuit Emphasis) and its variations [17-20]. Recent advancements in artificial intelligence even allow automating the generation of designs of new op amps [21, 22].

The goal of this work was to design and develop a low-power rail-to-rail operational amplifier that could operate on a single supply ranging from 5 to 30 V, or a dual supply of ±2.5 V and ±15 V. Key specifications included low voltage noise (< 4 µV, p-p from 0.1 to 10 Hz; ~ 13 nV/√Hz), ultralow input bias current (< 15 pA), and low offset voltage (< 500 µV). Similar performance is demonstrated by Analog Devices chip AD820 [23] (See Table 1). It is a well-known operational amplifier, appreciated for its versatility, low input bias current, and high common-mode rejection ratio, making it suitable for applications ranging from sensor signal conditioning to instrumentation. In comparison, popular op amp TL801 has low voltage noise (typical 9.2 µV, p-p from 0.1 to 10 Hz; ~ 37 nV/√Hz), low input bias current (< 120 pA), and low offset voltage (< 4 mV) on a single supply ranging from 4.5 to 40 V, or a dual supply of ±2.25 V and ±20 V [24]

Several modern operational amplifiers meet similar requirements to AD820 and could serve as alternatives. For instance, Analog Devices ADA4522-1 provides very low offset voltage (<5 µV), low input bias current (typical 50 pA), very low noise density (typical 5.8 nV/√Hz), operating with a single supply range of 4.5 V to 55 V or dual supply range ±2.25 V and ±27.5 V [25]. Additionally, the Maxim Integrated MAX44241 features low offset voltage (<5 µV), very low noise density (typical 9 nV/√Hz), operating on a single supply range from 2.7 V to 36 V or dual supply range ±1.35 V and ± 18 V [26]. The two formers are modern op amps with several parameters exceeding AD820, e.g. offset voltage and voltage noise, but demonstrate lower performance in input bias current values.

The design team at RD Alfa Microelectronics selected to adapt the AD820 prototype circuit to match available production capabilities, modify it as needed to address any shortcomings and leverage previous experience gained from developing the low-voltage four-channel amplifier chip aRD824 [27, 28]. ADA4522-1 was not selected as our prototype, as its circuitry and operating principle differ significantly from the standard approach used in most operational amplifiers (the ADA4522-1 uses a periodic offset voltage compensation principle), and we would need to spend considerably more time on research and development. Similar considerations were made to decline the prototyping of MAX44241. Additionally, it was assumed that our op-amp as an alternative to the AD820 would yield a higher consumer demand, and thus, our product would have larger commercial perspectives.

Table 1
**Comparison of selected parameters of various op amps similar to AD820**

| Parameter | aRD820 | AD820 [23] | TL801 [24] | ADA4522-1 [25] | MAX44241 [26] |
|---|---|---|---|---|---|
| Producer | RD Alfa Micro-electronics | Analog Devices | Texas Instruments | Analog Devices | Maxim Integrated/ Analog Devices |
| Supply Voltage Range | ±2.5 V to ±15 V (or single supply 5 to 30V) | ±2.5 V to ±15 V (or single supply 5 to 30V) | ±2.25V to ±20V (or single supply 4.5 to 40V) | ±2.25V to ±27.5V (or single supply 4.5 to 55 V) | ±1.35V to ±18V (or single supply 2.7 to 36 V) |
| Offset Voltage | < 500 µV | < 800 µV | < 4 mV | < 5 µV | < 5 µV |
| Input Bias Current | < 15 pA | < 25 pA | < 120 pA | typical 50 pA | < 600 pA |
| Voltage Noise, p-p, 0.1 -10 Hz | < 4 µV | typical 2 µV | typical 9.2 µV | typical 117 nV | typical 9 nV |
| Input Voltage Noise Density | > 13 nV/√Hz | typical 16 nV/√Hz (at 1 kHz) | typical 37 nV/√Hz (at 1 kHz) | typical 5.8 nV/√Hz | typical 9 nV/√Hz |

To evaluate our design, we constructed a prototype aRD820 and conducted a series of measurements to compare its performance against the standard AD820. In this paper, we detail these modifications and provide a report on our experimental setup and obtained results. Our results indicate that the modified operational amplifier aRD820 achieves similar, and in some aspects, advantageous performance characteristics when compared to the original AD820.

**Materials and methods**

The analysis of the AD820 operational amplifier and subsequent efforts to replicate it highlighted key challenges in matching the performance of the original chip. Initially, our production facility attempted to replicate the electrical scheme of the AD820, producing test chips that, unfortunately, exhibited performance shortcomings. A thorough investigation revealed several primary causes behind these deficiencies. First, the N-channel FET transistors used in our replication had lower characteristics compared to those used in the AD820, particularly in terms of noise performance. The noise voltage swing at the input was found to be between 5 µV and 10 µV, which was significantly higher than the performance of the AD820. Second, our transistors displayed a broader "Ohmic region" in their output voltage-current characteristics, resulting in lower transconductance in this region. When the input voltage approached the supply voltage limits, the gain of the amplifier decreased strongly, particularly in configurations using transistors with higher gate cutoff voltages. This phenomenon was due to the FET transistors entering the Ohmic region earlier than those in the AD820, thereby reducing the effective gain. Third, there was a significant spread in the gate cutoff voltages of our FET transistors, ranging from -0.5 V to -2.5 V. This variability complicated the process of setting a consistent reference current, leading to a lower yield of functional chips. The resistor values required to achieve the desired current varied significantly, exceeding the adjustment capabilities of the resistors available in our production technology. Fourth, the transistors exhibited an unacceptable increase in drain-gate reverse current at higher drain-gate voltages, which necessitated additional circuitry to limit the voltage across the input transistors. This issue further impacted the reliability of the operational amplifier at elevated supply voltages. Fifth, the thin-film resistors initially used in our design were found to be inadequate, with poor performance characteristics. We proposed replacing them with ion-implanted resistors, which provided a more consistent and reliable alternative for achieving the desired circuit stability.

Based on these findings, modifications were made to the electric schematics of the op-amp AD820 to adapt the design to the strengths and limitations of our production capabilities. We redesigned the input, second (pre-final), output, and current reference stages to enhance noise performance, expand the input signal range, stabilize the voltage across input transistors, and address phase stability concerns.

The input stage of the op-amp is crucial for initial amplification, typically implemented as a differential amplifier to enhance high input impedance and reject common-mode signals. Modifications were made to the input stage of the aRD820 relative to the AD820 to address three main issues: reducing noise, expanding the input range towards positive source rail +Vs, and stabilizing the voltage on the input transistors. By adding two source followers, noise reduction by a factor of five was achieved, and the allowable input signal range was increased by 0.4 V. The second (pre-final) stage of the aRD820 op-amp has a symmetrical structure, allowing equal signal propagation times in both the upper and lower arms, unlike the AD820, where the signal delay in the upper arm is compensated by capacitor. Calculations reveal that the aRD820's symmetrical design reduces signal delay and phase shift by 3.5 times compared to the AD820, even with lower operating currents. This design enhances the stability of the aRD820. Additionally, the aRD820 includes emitter followers to reduce offset voltage by ensuring consistent input currents for various transistor β values, further improving accuracy. In the output stage of aRD820, frequency compensation components - two capacitors and two resistors - were added to offset phase shifts introduced by earlier transistors in the input stage. Simulations showed that the aRD820's output stage has a lower gain compared to the AD820, but this was deemed acceptable due to sufficient gain margins. The aRD820's phase-frequency response was also improved, offering a phase margin of 89.6 degrees at 3 MHz, compared to 68.6 degrees for the AD820. This enhanced phase margin increases the aRD820's stability near the unity gain frequency. To improve yield in the aRD820 op-amp, the design eliminates the FET transistors that set the reference current in the AD820, replacing them with a simpler configuration. In the aRD820's reference source circuit, a single FET transistor is used only to initiate the circuit, and the reference current is set by the pnp transistor. Fine-tuning is achieved by trimming a series of resistors, ensuring that reference current drift remains within 5% to 10% across temperature and supply voltage variations.

Simulation of electric schematics of aRD820 realized in PSpice Microsim showed that its performance meets the requirements of this op amp and is close to the performance of aRD820. Thus, this schematic is a good example of how an electric circuit can be transformed to reduce its accent on high-quality N-channel FETs (in AD820) and achieve similar performance with npn and pnp transistors, average N-channel FETs and standard electric components like transistors and capacitors. This approach was previously used to create an electric scheme for four-channel rail-to-rail operational amplifier aRD824, based on an AD824 prototype [27].

In designing the op amp chip aRD820, a structured approach was undertaken to integrate three core transistor configurations: n-p-n, p-n-p, and n-FETs. The layered structural design, illustrated in Fig. 1, incorporates multiple semiconductor materials and doped regions that contribute to the overall performance of the amplifier. The Silica nitride layer and passivation layer are not shown there. This multi-layered configuration ensures the precise electrical characteristics required for the amplifier's operational stability and reliability. A topology of 18 layers was implemented for the aRD820 chip to achieve its complex architecture.

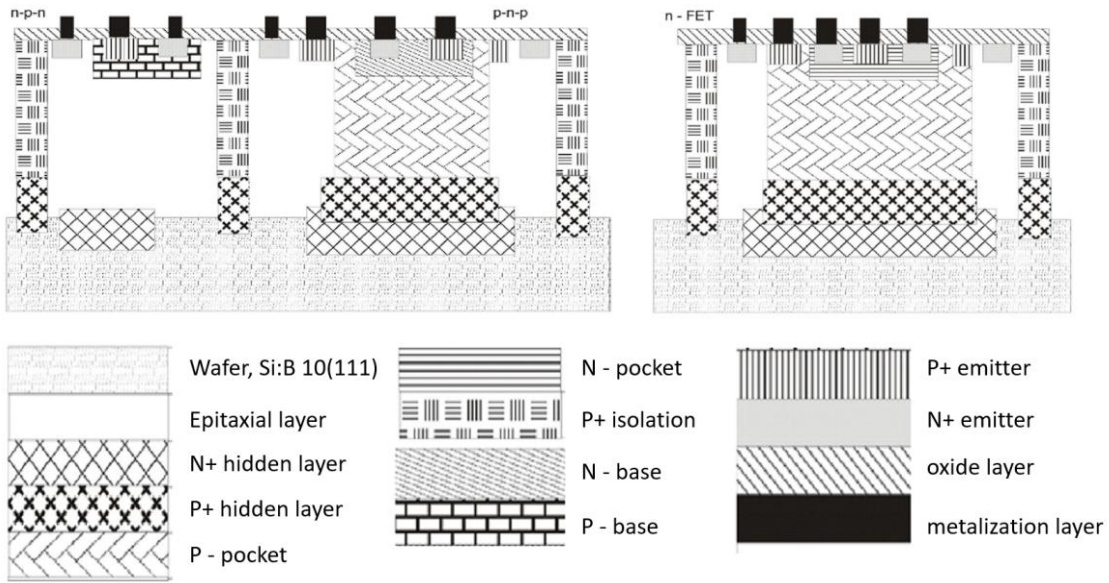

Fig. 1. **Structural scheme of the aRD820 operational amplifier**

All layers of the aRD820 were constructed on a boron-doped silicon Si (111) wafer, achieving p-type conductivity with a resistivity of 10 Ω·cm, which served as the foundational substrate for the device. The (111) crystal orientation of the wafer supports uniform layer deposition and enhances electrical properties. The initial epitaxial layer provides a high-quality, stable substrate that minimizes imperfections and enhances uniformity, essential for optimal chip performance. Building upon this, N+ and P+ hidden layers are included to isolate specific regions within the transistor, reducing leakage and ensuring controlled current flow. These layers play a key role in minimizing parasitic effects and allowing each transistor to function independently without interference. Next, the P-pocket and N-pocket regions are engineered to form the channels through which carriers flow in the transistors, optimizing carrier movement for the amplifier's responsiveness to input signals. P+ isolation layer distributes charge uniformly across the transistor's active areas, supporting consistent performance under varying conditions. The N-base and P-base layers form the core of the n-p-n and p-n-p transistors, defining their switching characteristics and enabling controlled current modulation crucial for desired amplifier response. P+ and N+ emitter layers serve as the emission points for holes and electrons to facilitate efficient carrier injection. The metallization layer provides the necessary interconnections, linking components within the op-amp and facilitating external connections. An oxide layer acts as an insulator, isolating active regions from the metallization layer and preventing electrical shorts.

Fig. 2 is a photo of the produced ap amp aRD820 chip. Fig 3. shows the schematic of connections of the aRD820 chip mounted on a TO-5 package.

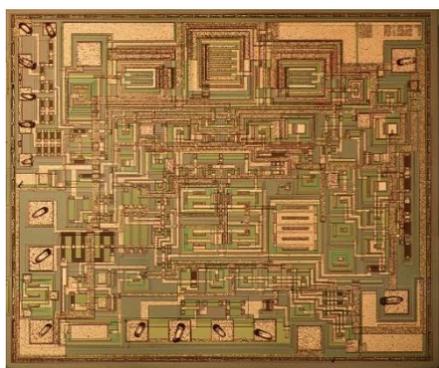

Fig. 2. **Photo of aRD820 operational amplifier chip**

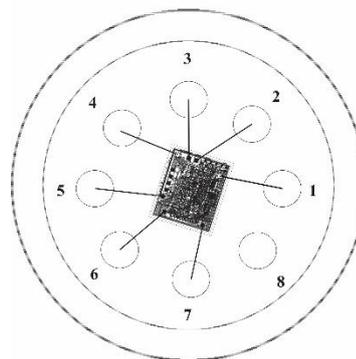

Fig. 3. **Schematics of connection of aRD820 operational amplifier chip on TO-5 package**

A module (specialized switching device) for testing aRD820 chips on wafers was developed and manufactured (Fig. 4). The module is inserted into a custom measurement system (Beta-210 + AVT-110), which ensures the measurement process is controlled by a computer. The measurement system sets the operating parameters (power supply voltage, input signals, time intervals, etc.), provides signal registration from the module, and adjusts the offset voltage. The module consists of a switching board and a probe head. On the switching board, relays are placed, which switch with the semiconductor wafer, as well as other elements that must be located in the immediate vicinity of the microchip crystals. On the probe head, needles are placed, which are connected to the corresponding areas of the semiconductor wafer. The measurement setup with the developed module ensures the following parameters can be set and measured: Offset Voltage, Input Bias Current, Input Offset Current, Open-Loop Gain, Common Mode Rejection Ratio, Power Supply Rejection Ratio, Output Saturation Voltage, Short-Circuit Current, and Quiescent Current.

A module for testing individual aRD820 chips in TO-5 packages was developed (Fig. 5). This module is intended to be used with the universal measuring device Beta-210. The module contains an electronics block similar to the wafer testing module. On the surface of the aRD820 microchip testing module, there is a contact socket designed for the easy connection of an aRD820 microchip. This module measures the same parameters as those listed for the wafer testing module.

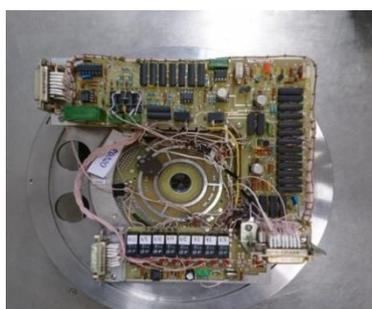

Fig. 4. **Module to inspect chips on a wafer**

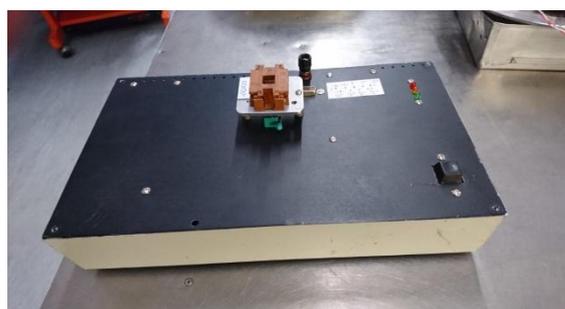

Fig. 5. **Module to test chips in TO-5 packages**

Measurement results are provided in detail in the following section.

**Results and discussion**

aRD820 chips were produced, and subsequent tests were made on aRD820 chips in both wafer form and within TO-5 packages. In wafer form, tests were made mainly for classification purposes – to distinguish good and damaged chips. Good chips were inserted in TO-5 packages and tested extensively.

Table 2 summarizes data on the planned characteristics of aRD820 and test results from 31 selected chips obtained from one wafer, when supply voltage ±2.5 V is applied. Measurements were made at normal temperature ($T_{NORM}$ = 25 ºC) and extreme temperatures - low ($T_{MIN}$ = -55 ºC) and high ($T_{MAX}$ = 125 ºC). The offset voltage for aRD820 was planned to be within range -0.5 … 0.5 mV at normal temperatures and -1.5… 1.5 mV at extreme temperatures. Tests show that this result is achieved. In extreme temperatures measured data is even within smaller limits of -0.51… 0.55 mV. In comparison, AD820 data sheet [23] shows similar results – max 0.8 mV in normal temperatures and 1.2 mV at extreme temperatures.

Table 2

**Planned and measured parameters of aRD820 chips when the supply voltage is ±2.5V. Measurements were made on 31 chips in TO-5 packages**

| No. | Parameter | Temperature | aRD820 (planned) | | aRD820 (measured) | | | Units |
|---|---|---|---|---|---|---|---|---|
| | | | Min | Max | Min | Average | Max | |
| 1 | Offset Voltage | $T_{NORM}$ = 25 ºC | -0.5 | 0.5 | -0.44 | -0.06 | 0.46 | mV |
| | | $T_{MIN}$ = -55 ºC | -1.5 | 1.5 | -0.51 | -0.06 | 0.47 | |
| | | $T_{MAX}$ = 125 ºC | -1.5 | 1.5 | -0.51 | -0.07 | 0.55 | |
| 2 | Input Bias Current (+) | $T_{NORM}$ | -15 | 15 | -2.34 | 0.26 | 1.32 | pA |
| | | $T_{MIN}$ | -4000 | 4000 | -498.57 | 69.23 | 300.45 | |
| | | $T_{MAX}$ | -4000 | 4000 | -510.57 | 47.23 | 288.45 | |
| 3 | Input Offset Current | $T_{NORM}$ | -10 | 10 | 3.70 | 4.54 | 5.28 | pA |
| | | $T_{MIN}$ | -500 | 500 | 111.13 | 136.60 | 159.58 | |
| | | $T_{MAX}$ | -500 | 500 | 148.17 | 182.14 | 212.77 | |
| 4 | Open-loop gain $R_L$ = 2 kΩ | $T_{NORM}$ | 10 | | 64.51 | 262.68 | 1338.71 | V/mV |
| | | $T_{MIN}$ | 20 | | 70.52 | 286.27 | 1463.43 | |
| | | $T_{MAX}$ | 20 | | 47.01 | 190.85 | 975.62 | |
| | $R_L$ = 10 kΩ | $T_{NORM}$ | 50 | | 354.57 | 880.01 | 5192.02 | |
| | | $T_{MIN}$ | 80 | | 387.60 | 963.92 | 5675.74 | |
| | | $T_{MAX}$ | 80 | | 258.40 | 642.61 | 3783.83 | |
| | $R_L$ = 100 kΩ | $T_{NORM}$ | 250 | | 944.47 | 1866.02 | 6812.22 | |
| | | $T_{MIN}$ | 400 | | 688.31 | 1367.24 | 4964.60 | |
| 5 | Output Saturation Voltage (High) $I_{SOURCE}$ = 20 µA | $T_{NORM}$ | | 14 | 2.85 | 2.92 | 3.14 | mV |
| | | $T_{MIN}$ | | 20 | 2.27 | 2.34 | 2.52 | |
| | | $T_{MAX}$ | | 20 | 2.06 | 2.13 | 2.29 | |
| | $I_{SOURCE}$ = 2 mA | $T_{NORM}$ | | 110 | 101.10 | 104.30 | 107.53 | |
| | | $T_{MIN}$ | | 160 | 147.10 | 151.86 | 156.46 | |
| | | $T_{MAX}$ | | 160 | 98.07 | 101.24 | 104.31 | |
| | $I_{SOURCE}$ = 15 mA | $T_{NORM}$ | | 1500 | 1083.31 | 1118.78 | 1164.13 | |
| | | $T_{MIN}$ | | 1500 | 1184.24 | 1223.87 | 1272.59 | |
| | | $T_{MAX}$ | | 1500 | 789.49 | 815.91 | 848.39 | |
| 6 | Output Saturation Voltage (Low) $I_{SINK}$ = 20 µA | $T_{NORM}$ | | 7 | 5.71 | 5.85 | 6.28 | mV |
| | | $T_{MIN}$ | | 10 | 3.69 | 5.41 | 7.13 | |
| | | $T_{MAX}$ | | 10 | 4.06 | 5.96 | 7.84 | |
| | $I_{SINK}$ = 2 mA | $T_{NORM}$ | | 55 | 31.85 | 37.09 | 53.64 | |
| | | $T_{MIN}$ | | 80 | 21.10 | 24.42 | 35.54 | |
| | | $T_{MAX}$ | | 80 | 23.21 | 26.87 | 39.09 | |
| | $I_{SINK}$ = 15 mA | $T_{NORM}$ | | 500 | 319.57 | 346.56 | 460.77 | |
| | | $T_{MIN}$ | | 500 | 211.72 | 237.85 | 367.72 | |
| | | $T_{MAX}$ | | 500 | 232.90 | 261.63 | 404.50 | |
| 7 | Voltage noise, 0.1 Hz to 10 Hz | $T_{NORM}$ | | 4 | 2.15 | 3.05 | 3.73 | µV, p-p |
| 8 | Voltage noise density | $T_{NORM}$ | 13 | | 13.50 | 15.17 | 18.13 | nV/√Hz |

The planned input bias current for aRD820 chip is below the absolute value of 15 pA at normal temperatures and below 4 nA at extreme temperatures. Tests show that at normal temperatures the absolute value is below 2.34 pA and 0.511 nA for extreme temperatures, thus fitting expected limits. In comparison, the input bias current for AD820 is below 25 pA for normal temperature and below 5nA at extreme temperatures.

The planned input offset current for aRD820 chip is below the absolute value of 10 pA at normal temperatures and below 500 pA at extreme temperatures. Tests show that at normal temperatures the absolute value is below 5.28 pA and below 213 pA for extreme temperatures, thus fitting expected limits. In comparison, the input offset current for AD820 is below 20 pA for normal temperature and is typically 500 pA at extreme temperatures. Fig. 4 shows a histogram of the input offset current for aRD820 at normal temperatures. These data can be well-fitted with normal distribution.

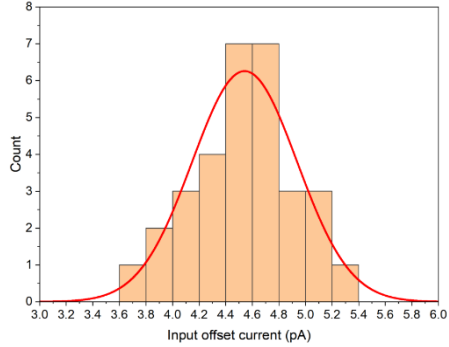

Fig. 6. **Histogram of Input offset current for aRD820.** 31 chips explored, source voltage Vs = ±2.5V, temperature $T_{NORM}$ = 25°C. Fitting is made by normal function

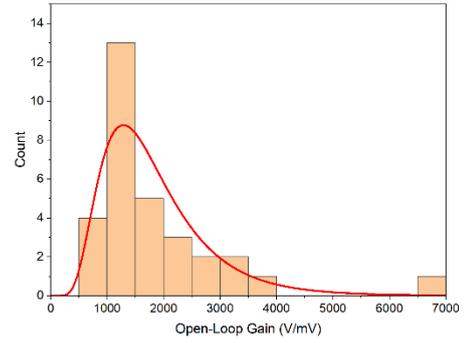

Fig. 7. **Histogram of Open-Loop Gain for aRD820.** 31 chips explored, source voltage Vs = ±2.5V, temperature $T_{NORM}$ = 25°C, load resistance $R_L$ = 2 kΩ. Fitting is made by lognormal function

The planned open-loop gain for aRD820 chip at load resistance 2 kΩ is above 10 V/mV at normal temperatures and above 20 V/mV at extreme temperatures. Tests show that this signal is above 64 V/mV at normal temperatures and above 47 V/mV at extreme temperatures. Fig. 7 shows a histogram of open-loop gain at normal temperature and load resistance 2 kΩ. Good fitting for this histogram is made by lognormal function. Data for load resistances 10 kΩ and 100 kΩ mA are given in Table 2. The tests show that planned data values are met. In comparison, AD820 shows similar performance. An extended view of open-loop gain vs load resistance for aRD820 and AD820 is given in Fig. 8. It shows data for aRD820 at normal temperature when the source voltage Vs = ±2.5 V and ±15 V, and data for AD820 at normal temperature when the source voltage Vs = 0, 5 V and ±15 V. Open-loop gain values at Vs = ±2.5 V of aRD820 can be compared to values of Vs = 0, 5V for AD820. It is seen that in this case aRD820 shows higher gain for about 2 times at 2 kΩ, about 4 times higher gain at 10 kΩ, and a bit higher gain at 100 kΩ. Similarly, values for open-loop gain at Vs = ±15V of aRD820 and AD820 can be compared. In this case, values are very close in the range 2 - 10 kΩ, but close to 100 kΩ AD820 outperformes aRD820 by a factor of 2. Figure 8 also indicates that the open-loop gain versus load resistance, shown on a logarithmic scale, exhibits linearity above 1 kΩ for the AD820. In contrast, strict linearity is not observed for the aRD820 in the same region. If data points for aRD820 will increase, probably, at least some region on linear dependence will appear. For both op amps, their curves for low and high source voltages show similarity in shape. Another notable observation is that for AD820 at higher source voltages open-loop gain is higher. In contrast, for aRD820 this dependence is the opposite. The likely reason for this abnormal dependence in aRD820 is the increased measurement error of Vs at ±2.5V (the error can reach up to 40%). This occurs because the open-loop gain measurement at Vs = ±2.5 V is conducted at input voltages that are five times lower than those for the measurement at ±15 V. To clarify, let's assume the op amp under test has a gain of 1,000,000. In that case, the voltage between the op amp's inputs will be 4 µV at ±2.5V and 20 µV at ±15V. Measuring values as small as 4 µV and 20 µV is essentially the task. Achieving low measurement error at levels around 2 µV to 4 µV is extremely challenging. Therefore, in our measurements, we focus on determining "pass" or "fail" while allowing for an error margin of approximately 40%. Thus, the actual open-loop gain value at ±2.5 V supply may differ from the value shown in Figure 8.

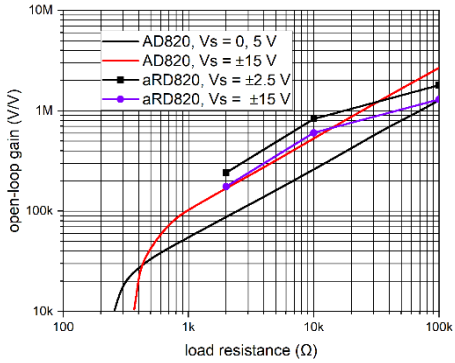 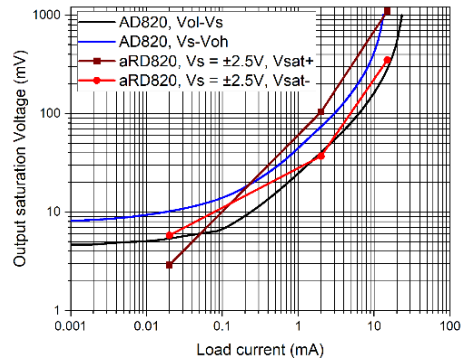

Fig. 8. **Open-loop gain vs. Load Resistance at normal temperature for aRD820 at source voltage ±2.5 V and ±15V, and for AD820 at source voltage 0, 5 V and ± 15V**

Fig. 9. **Output saturation voltage for positive and negative rails vs load current for aRD820 at source voltage ±2.5V, and for AD820**

The output saturation voltage (high) is the difference between the highest possible output voltage and the positive supply voltage. The planned and measured data of output saturation voltage (high) of aRD820 for source current values of 0.02, 2, and 15 mA are given in Table 2. Experimental results confirm that the saturation voltage is below the planned maximal limits. Additionally, these values are much similar to AD820. The output saturation voltage (low) is the difference between the lowest possible output voltage and the negative supply rail. The planned and measured data of output saturation voltage (low) of aRD820 for sink current values of 0.02, 2, and 15 mA are also given in Table 2. Experimental results confirm that the saturation voltage remains below the specified maximum limits. These values are lower than the corresponding output saturation voltages (high). This dependence is also seen in AD820 [23]. Fig. 9 shows output saturation voltage dependence from load current. Output saturation voltage (high) in this graph corresponds to "Vsat+" values of aRD820 and "Vs-Voh" values of AD820. Output saturation voltage (low) in this graph corresponds to "Vsat-" values of aRD820 and "Vol-Vs" values of AD820. We see that data point of "Vsat-" strictly fits on "Vol-Vs" showing good correspondence between aRD820 and AD820. For output saturation voltage (high) good fit is seen for high load current above 3 mV, but for lower load, aRD820 show lower values compared to AD820.

Table 2 shows the planned and measured values of voltage noise of aRD820 at 0.1 Hz to 10 Hz. Planned values are below 4 µV peak to peak (p-p), and measured values were below this limit and on average 3 µV, p-p. Table 1 shows corresponding values for several alternative op amps, e.g., for AD820 typical voltage noise is 2 µV and 9.2 µV for TL801. These tables show also input voltage noise density values. For aRD820, planned values are above 13 nV/√Hz, measured values are 13.5 - 18.1 nV/√Hz. For AD820, typical values for input voltage noise density is 37 nV/√Hz.

Fig. 10 shows the relationship between open-loop gain and frequency for aRD820 and AD820. In log scale the gain drops linearly with the increase of frequency giving the open-loop gain of about 100 dB at 20 Hz and 0 dB at about 1.5 - 2 MHz. This shows that aRD820 has achieved much similar performance to AD820.

Fig. 11 shows the input bias current vs temperature for aRD820 and AD820. At 60 ºC input bias current for both op amps overlap reaching about 20 pA. Below 60 ºC, aRD820 shows higher input bias current than AD820, saturating to about 10 pA below 20 ºC. Above 60 ºC, aRD820 shows lower input bias current than AD820.

Fig. 12-13 show the small signal response (Unity-Gain Follower) to a time-varying signal — test results for aRD820 and AD820. The input signal was square-shaped. Both op amps demonstrate similar performance: aRD820 reaches a signal increase of approximately 40 mV in 0.1 µs, and the altered signal stabilizes in approximately 1 µs, whereas the AD820 microchip achieves a signal increase of 40 mV in approximately 0.2 µs, with the altered signal stabilizing in about 0.5 µs.

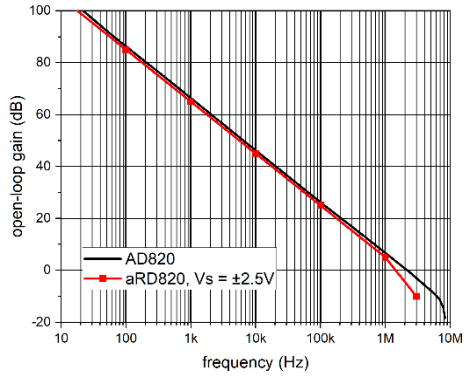

Fig. 10. **Open-loop gain versus frequency for aRD820 at source voltage ±2.5 V and AD820**

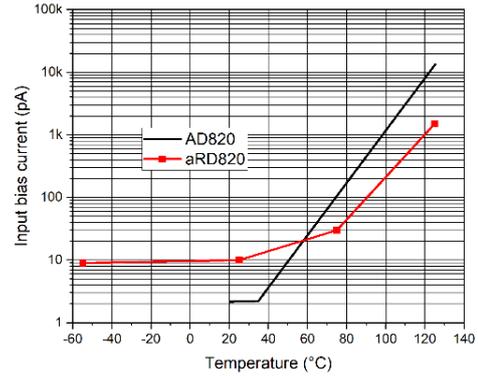

Fig. 11. **Input bias current dependence on temperature for aRD820 and AD820**

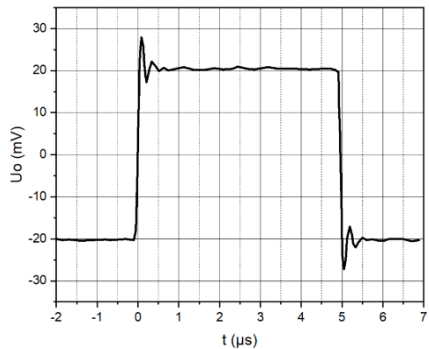

Fig. 12. **Unity-Gain Follower dependence in time for aRD820 at source voltage ± 15 V and load resistance 10 kΩ**

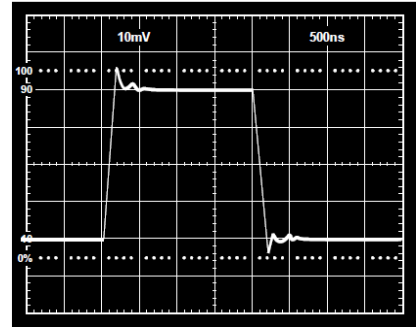

Fig. 13. **Unity-Gain Follower dependence in time for AD820 [23]**

## Conclusions

The development and testing of the aRD820 operational amplifier present an advancement in low-power, rail-to-rail op-amp technology. By adapting and modifying the design of the well-known AD820 to accommodate the manufacturing capabilities and material constraints of RD Alfa Microelectronics, the aRD820 achieves comparable performance characteristics while addressing specific limitations encountered in the AD820 replication process.

The experimental results of the aRD820 demonstrate that it meets or exceeds planned performance benchmarks across several key parameters. Tests show that its open-loop gain, input bias current, input offset current, and output saturation voltages are well within desired limits. The aRD820 provides an open-loop gain of approximately 100 dB at low frequencies and maintains stability over a broad range of load resistances and supply voltages. Additionally, voltage noise characteristics fall within planned specifications, ensuring the amplifier's suitability for sensitive signal applications.

Overall, the aRD820 successfully replicates the core functionalities of the AD820 with enhancements tailored for RD Alfa Microelectronics' manufacturing processes, offering comparable performance in an economically feasible package. The results validate the design modifications and provide a solid foundation for future improvements and commercial applications.

## Acknowledgements

Research activities were funded by Latvian Recovery and Resilience Mechanism Plan under reform and investment direction 5.1.r. "Increasing productivity through increasing the amount of investment in R&D" (No. 5.1.1.2.i.0/1/22/A/CFLA/002).

## Author contributions:

Conceptualization, D.K., S.R., and M.L.; data curation, D.K. and S.R.; formal analysis, D.K., S. R., M.L., and A.A.; funding acquisition, M.L.; investigation, D.K., S.R., M.L., and A.A.; methodology, D.

K., S.R., and M. L.; software, D.K.; supervision, S.R., validation, S.R. and M.L.; visualization, D.K. and A.A.; writing - original draft preparation, D.K. and A.A.; writing - review and editing, D.K., A.A., S.R., and M.L. All authors have read and agreed to the published version of the manuscript.